# Low-loss Tunable 1-D ITO-slot Photonic Crystal Nanobeam Cavity

Rubab Amin[1], Mohammad H. Tahersima[1], Zhizhen Ma[1], Can Suer[1], Ke Liu[1,2], Hamed Dalir[3], Volker J. Sorger[1,*]

[1]Department of Electrical and Computer Engineering, George Washington University,
800 22nd St., Science & Engineering Hall, Washington, DC 20052, USA
[2]Key Laboratory of Optoelectronics Technology, Ministry of Education, Faculty of Information Technology,
Beijing University of Technology, Beijing 100124, China
[3]Omega Optics Inc., 8500 Shoal Creek Blvd., Bldg. 4, Suite 200, Austin, TX 78757, USA
*Corresponding Author E-mail: *sorger@gwu.edu*

**Tunable optical material properties enable novel applications in both versatile metamaterials and photonic components including optical sources and modulators. Transparent conductive oxides (TCOs) are able to highly tune their optical properties with applied bias via altering their free carrier concentration and hence plasma dispersion. The TCO material indium tin oxide (ITO) exhibits unity-strong index changes, and epsilon-near-zero behavior. However, with such tuning the corresponding high optical losses, originating from the fundamental Kramers-Kronig relations, result in low cavity finesse. However, achieving efficient tuning in ITO-cavities without using light matter interaction enhancement techniques such as polaritonic modes, which are inherently lossy, is a challenge. Here we discuss a novel one-dimensional photonic crystal nanobeam cavity to deliver a cavity system offering a wide range of resonance tuning range, while preserving physical compact footprints. We show that a vertical Silicon-slot waveguide incorporating an actively gated-ITO layer delivers ~3.4 nm of tuning. By deploying distributed feedback, we are able to keep the *Q*-factor moderately high with tuning. Combining this with the sub-diffraction limited mode volume (0.1 ($\lambda/2n$)$^3$) from the photonic (non-plasmonic) slot waveguide, facilitates a high Purcell factor exceeding one thousand. This strong light-matter-interaction shows that reducing the mode volume of a cavity outweighs reducing the losses in diffraction limited modal cavities such as those from bulk $Si_3N_4$. These tunable cavities enable future modulators and optical sources such as tunable lasers.**

1. Introduction

Tunable and compact on-chip components such as light sources, electro-optic modulators, and detectors are desired for optical interconnects and photonic circuitry [1,2]. Large difference in characteristic wavelength of light and electrons lead to an inherently weak interaction between light and matter, resulting in bulky and inefficient optoelectronic components. A well-known solution to reduce the device footprint on photonic chips is the use of optical cavities to enhance light-matter interaction by reducing modal volume or increasing the quality factor of the optical effect. One dimensional (1D) photonic crystal nanobeam (PCNB) structure offers strong light confinement in the cavity formed by a periodic array of air holes along the waveguide propagating direction, which takes advantage of its relatively high-quality (*Q*) factor in a small modal volume approaching the diffracting limit of light (i.e., high Purcell effect). For optical tunability and reconfigurability there are a variety of options such as modifying coupling, scattering, light propagation direction, or index changes of the optical mode. The latter can be achieved via altering the optical index of an 'active' material electro-optically, opto-mechanically, thermo-optically, or all-optically. In terms of material choices for electro-optic tuning, Indium Tin Oxide (ITO) is a promising material, since it is a compatible material with electronic circuitry and shows unity-strong gate tunable change in its refractive index at telecom bands. Typical mechanisms are depletion and accumulation of carriers when electrostatically gated. Recent reports of tunable resonant plasmonic metasurfaces using ITO have demonstrated the opportunity to realize thin metafilm absorbers and for beam steering applications [3-5]. ITO-based on-chip modulators are also being explored by electrically induced change in the free carrier concentration of ITO overlapping with the propagating optical mode. [6, 7].

A photonic platform capable of shifting the resonance frequency of a high-*Q* cavity PCNB is of interest for photonics, because the large *Q*-factor increases the sensitivity to an optical reconfigurable active layer useful for tunable lasers, optical filters, or electro-optic modulators. Here, we investigate the design of a 1D PCNB integrated with the active ITO material by employing a vertical slot structure. Such structure can highly confine an optical mode in the active region to enhance the light-matter interaction, and hence, effective tunability, of ITO for a wide range of active on-chip components; such as lasers and optical modulators. In the following discussion, the nomenclature of the states of operation for the proposed device relate to the optical transmission characteristics through the cavity, rather than the applied bias; we use the device ON state to imply high optical transmission whereas the OFF state signifies the lossy state corresponding to a drop in the transmission.

## 2. Device Design

2.1. ITO Slot Waveguide

We employ a vertical Silicon slot waveguide with a thin ITO film in the dielectric slot gap region. This modal configuration allows the slot waveguide mode to be tightly confined to the active (ITO) region. In fact, such semiconductor-insulator-semiconductor (SIS) mode is able to allow sub-diffraction limited optical modes despite the absence of metals; here the jump of the dielectric permittivity across the high index silicon claddings and the low index center section (comprised of oxide/ITO/oxide) allows for optical confinement in the SIS configuration (Fig. 1). With a certain bias voltage, both the propagation constant and absorption coefficient of the fundamental transverse magnetic (TM)-like mode change considerably [8]. The SIS slot waveguide uses a 10 nm ITO thin film layer sandwiched between two p-Si layers of 150 nm separated by two oxide layers each 5 nm to facilitate gating (Fig. 1). This ITO layer can be considered as a degenerate semiconductor with a large number of electrons. For the oxides, we selected the high-k dielectric $HfO_2$ for the following two reasons: a) $HfO_2$ has a high static permittivity ($\epsilon_{HfO_2} \sim 25$) which increases the electrical capacitance, and thus allows lowering the gating voltage in tuning this cavity, thus helping to lower the energy overhead of active reconfiguration [9]; and b) $HfO_2$ thin films can be reliably grown with the atomic layer deposition (ALD) process [10]. With ITO biased as one electrode of a capacitor, it can achieve the three known states of accumulation, depletion, and inversion. For example, in accumulation, free carriers are accumulated in the interface of the ITO and the oxide, thus changing the carrier concentration via the Drude model. The optical property of the active material therefore changes, resulting in strong optical tunability of the cavity as discussed in this work [11]. We assume that the external electrodes form ohmic contacts with ITO and p-type Si. In reality, this may be not straight forward to achieve, but work-function engineering or partial selective contact area treatments could be used. For instance, we previously developed a process that resulted in lowering the contact resistance of ITO by 2-orders of magnitude with an oxygen plasma treatment [6]. When a negative voltage bias is applied to the ITO gate, free carriers are accumulated at the interface between the dielectric and ITO layer, which alters the optical mode's index and subsequently that of the optical cavity. Compared to previous work on a single metal-oxide-semiconductor (MOS) structure [12], the proposed design consists of two SIS stacks connected in parallel to each other (treating ITO as a highly-doped semiconductor). Since two electron-accumulated layers are formed at both dielectric-ITO interfaces, the thickness of the activated ITO screening by sweeping in carriers corresponding to the electro-optic change in the material dictated by the Thomas-Fermi screening length, $\lambda_{TF} \sim 1\ nm$ becomes twice; i.e. ~2 nm for both sides combined. As a result, the electro-optic effect increases considerably. In praxis, a 1/e decay length of about 5 nm was measured before [13], and high index tunability has been experimentally verified over $1/e^2$ (~10 nm) thick films [2,6]. Also the improved electrostatics from double gating design allows improved tuning efficiency and the corresponding gradual index change into the ITO layer from the oxide interface is smoother than that arising from one sided gating.

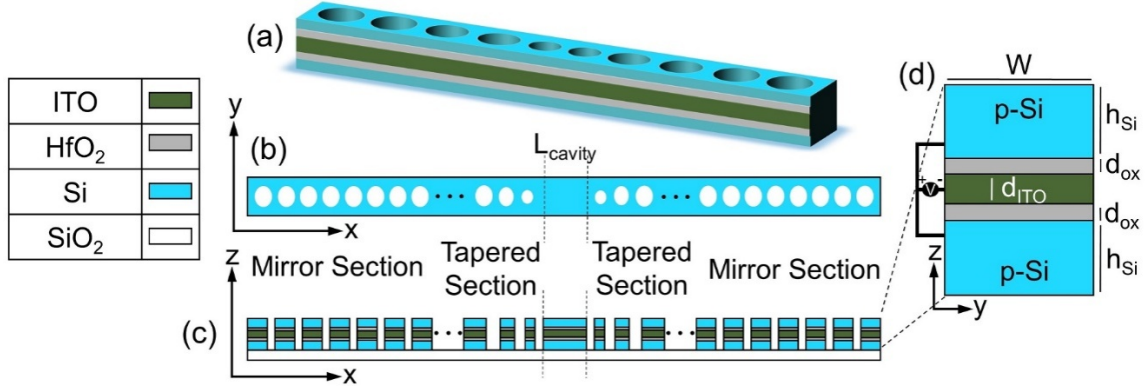

**Figure 1:** Schematic of the proposed 1D vertical slot ITO PCNB cavity with air holes. (a) 3D perspective view, (b) Top view, and (c) Side view (longitudinal cross-section) showing the mirror section and tapered section holes forming the Fabry-Pérot-like cavity; cavity length, $L_{cavity}$ = 100 nm, hole period of mirror section, a = 365 nm; minimum hole distance of taper section, $a_{min\_taper}$ = 230 nm; hole radius, r = 0.28a; number of taper hole pairs, n = 14; number of mirror hole pairs, m = 18; waveguide height, h = 320 nm. (d) The cross-sectional mode structure with the ITO thin film sandwiched between two p-Si claddings with oxides on either side for electrostatic gating. The relevant parameters are: W = 300 nm, $h_{Si}$ = 150 nm, $d_{ox}$ = 5 nm, and $d_{ITO}$ = 10 nm.

## 2.2. 1-D Photonic Crystal Nanobeam Cavity

Photonic crystals (PhC) are a class of optical media represented by natural or artificial structures with periodic modulation of refractive indices. A 1D photonic crystal nanobeam cavity (PCNB) incorporated with the ITO vertical slot waveguide is formed by creating vertical openings forming mirror and taper sections [14]. The nanobeam cavity considered here is essentially a wavelength-scale Fabry-Pérot etalon formed by sandwiching a defect section between 1D photonic crystal Bloch mirrors and tapered sections [15]. Light in the longitudinal direction in the nanobeam is confined by a principle of total internal reflection (Fig. 1a). Similar to a Fabry-Pérot spacer, the mirror and tapered section holes at the center of the PhC effectively confine the light in the propagating direction [16]. In other words, the taper sections are only used for improving mode impedance mismatch and minimizing the reflection. The design parameters of the 1D PCNB cavity include the cavity length of 100 nm; hole period of mirror section, a is 365 nm; minimum hole distance of taper section, $a_{min\_taper}$ is 230 nm; hole radius, r = 0.28a; number of taper hole pairs, n = 14; number of mirror hole pairs, m = 18; waveguide height, h is 320 nm. This 1D photonic crystal is simulated by the excitation of a fundamental TM-like 1st order dipole source from the z-axis and the length span of the device is in the x-direction. The length and width of the PCNB structure are 23 $\mu$m and 300 nm, respectively (Fig. 1b and d). The PCNB structure is integrated into a typical SOI substrate with a buried oxide of 1 $\mu$m thickness.

## 3. Material/Modal Tuning Aspects

The active material changes its optical properties with tuning based on the established models. For ITO, we characterize the corresponding complex refractive index variation from electrical tuning defined by the Drude model. The permittivity $\tilde{\epsilon} = \epsilon' - i\epsilon''$, is given by

$$\tilde{n}^2 = \tilde{\epsilon} = \epsilon_\infty - \frac{\omega_p^2}{\omega(\omega+i\gamma)} \qquad (1)$$

where $\epsilon_\infty$ is the high-frequency dielectric constant or background permittivity, $\omega$ is the angular frequency of the illuminating light, $\gamma = 1/\tau$ is the carrier scattering rate (i.e. collision frequency), and $\omega_p$ is the unscreened plasma angular frequency defined by $\omega_p^2 = N_c q^2/\epsilon_0 m^*$ [17,18]. Here $N_c$ is the gate-tunable carrier concentration, $q$ the electronic charge, $\epsilon_0$ the permittivity of vacuum, and $m^*$ is the conductivity effective carrier mass. The conductivity effective mass, $m^*$ is taken as $0.35m_0$, where $m_0$ is the free electron mass [6,19]. As the drive bias is increased, free carriers leading to dispersive effects are created and a corresponding net increase in the carrier concentration occurs. Unlike doped silicon, ITO whose chemical composition is usually provided by suppliers as $In_2O_3:SnO_2$, can be considered an alloy as the concentration of Sn relative to indium can be as high as 10%. In the near-infrared, ITO is quasi-metallic since its free electrons dictate its optical response. In fact, ITO and related transparent conducting oxides (TCOs) have recently been explored as plasmonic materials in the optical frequency range [8,20-24]. The Drude model characterizes the ITO material accurately within our specific wavelengths of interest (i.e., NIR regime) [18]. Several previous studies have calculated the permittivity of ITO using the experimentally measured reflectance and transmittance, and here a fitting result of Michelotti et. al is choosen whereas $\epsilon_\infty$ and $\gamma$ depend on the deposition conditions. In our analysis, we use $\epsilon_\infty = 3.9$ and $\gamma = 1.8 \times 10^{14}$ rad s$^{-1}$, respectively [18,24]. The plasma frequency, $\omega_p$ is determined by the carrier density $N_c$, which is equal to the concentration of Sn atoms, and thus can be within the range of $10^{19}$–$10^{21}$ cm$^{-3}$ depending on the deposition conditions, defect states, and film thicknesses [8,20]. The carrier concentration levels, effective indices and extinction coefficients corresponding to the states of operation in this work are listed in table 1.

**Table 1:** ITO carrier concentration levels, effective indices and extinction coefficients corresponding to the states of operation for the tunable cavity.

| States of operation | ITO Carrier Concentration (cm$^{-3}$) | Modal effective index, $\tilde{n}_{eff}$ | |
|---|---|---|---|
| | | Real part, $n_{eff}$ | Imaginary part, $\kappa_{eff}$ |
| ON | $10^{19}$ | 2.27 | $2.14 \times 10^{-4}$ |
| OFF | $10^{20}$ | 2.25 | $2.53 \times 10^{-3}$ |

While ITO is able to tune its index via the carrier-dependent Drude model (similar to Silicon), yet ITO exhibits a number of significant advantages over Si with respect to the index tuning; firstly, the carrier concentration in ITO can exceed that of Si by at least an order of magnitude; the concentration of indium atoms in ITO reaches a few percent, which is significantly above attainable donor concentration in Si [25]. Secondly, the effect of changing the carrier concentration in ITO on its refractive index is more dramatic than in Silicon. This can be attributed to the higher bandgap and consequently lower refractive index of ITO compared to that of Silicon. If the change of the carrier concentration $\delta N_c$ (e.g. due to an applied bias) causes a change in the relative permittivity (dielectric constant) $\delta\epsilon$, the corresponding change in the refractive index can be written as $\delta n = \delta\epsilon^{1/2} \sim \delta\epsilon/2\epsilon^{1/2}$, hence the refractive index change is greatly enhanced when the permittivity $\epsilon$ is small [11]. The effect of tuning from ON to OFF state on the material optical properties such as the complex refractive index and complex permittivity are depicted in Fig. 2(a,b).

The modal field for the ON- state, is strongly confined in the active region of ITO layer, where most of the field is confined in the ITO with both oxide layers of ~20 nm (Fig.2c). By means of the effective optical capacitance in the gap (oxide + ITO), arising from the continuity of the displacement field at the material interfaces, the light is strongly confined in the gap region in the vertical slot structure (Fig. 2c) [26]. The ITO index is lower than that of the high-k oxide and the cladding p-Si layers. Light being prone to occupy the spaces with lower index in such a slot – it is apparent that the light tends to be squeezed predominantly into the active ITO layer. The mirror symmetry in the vertical direction also helps the index contrast from the p-Si cladding layers to facilitate confinement into the active ITO. Upon gating, the OFF-state modal field illumination pattern does not observe drastic change as we tune carrier concentration by only an order of magnitude, so that the material is basically still in the n–dominant region and far from the epsilon-near-zero (ENZ) region [11,25]. It can be noticed, that the mode becomes more confined with tuning as the modal field is squeezed into the active region (i.e. ITO), which is consistent with our previous findings [11,25]; as the activated ITO approaches an ENZ (~6-7×10$^{20}$ cm$^{-3}$) condition, the modal confinement increases. The effective index also follows the ITO material characteristics due to the increased confinement, i.e. $n_{eff}$ decreases and $\kappa_{eff}$ increases with tuning.

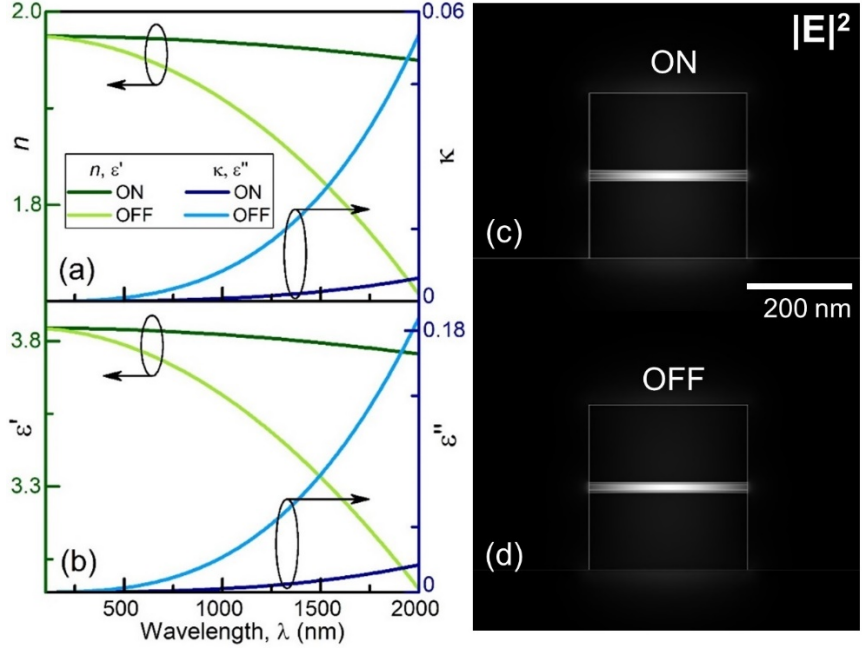

**Figure 2:** (a) The complex index of refraction (n and $\kappa$), and (b) the complex permittivity ($\epsilon'$ and $\epsilon''$) for ITO. The modal E-field intensity profiles for (c) ON and (d) OFF states are shown at the cross-sectional mode structure. The magnitude of the E-field, $|E|^2$ is plotted for the mode profiles.

## 4. Results and Discussions

The cavity mode is excited by placing an electric dipole source with z-orientation inside the vertical slot cavity. The 1-D ITO PCNB cavity exhibits a TM-like fundamental 1st order mode (i.e., dominant E-field in the z-direction) [Fig. 2c]. The electric field of the TM mode extends longitudinally along the waveguide, which is consistent with free-standing dual-polarized silicon PhC nanobeam cavities [27]. In addition, the electric field profile is confined in the low index active region between these air holes. As expected, light-matter interaction (LMI) is enhanced in this type of structure compared to a bulk photonic structure due to increased confinement. The feedback from the mirror sections and subsequent impedance matching from the tapered sections allow the confinement of light inside the cavity (centered at x = 0) in the longitudinal propagating direction (x-orientation), which is evident in the cavity profiles for the xy-plane and xz-plane (Fig. 3(a,b)). The cavity as a means of storage for the energy (i.e. $|E^2|$) is also discernible from the cavity xy and xz profiles as both exhibit the highest field intensity near x = 0 (cavity region).

Tuning is an effect of change in the effective refractive index, $n_{eff}$. The subsequent broadening of the cavity resonances with respect to wavelength can be related back to the loss in the modal absorption, corresponding to the effective extinction coefficient, $\kappa_{eff}$. The shift in resonance with the change in carrier concentration, i.e. tuning, $\Delta \lambda$, results from the change in the effective refractive index (real part), $\Delta n_{eff}$, from the aforementioned modal tuning properties. Our results show a considerable amount of wavelength tuning, $\Delta \lambda \sim 3.4$ nm, for the slot ITO PCNB cavity (Fig.3c). From ON to OFF state, the cavity exhibits blue-shifts in the resonances originating from the effective index decrease. Loss also increases with tuning as a direct result from the Kramers–Kronig relations. Regarding the tuning range (i.e., carrier concentration levels), we have managed to keep this and subsequent broadening of the resonances low due to: a) selectively reducing the allowable carrier concentration tuning range in the n–dominant region away from the ENZ region, and b) by selecting a low carrier concentration initial point (ON state) of $10^{19}$ cm$^{-3}$, where ITO behaves as a dielectric. As the resonances broaden with tuning, the cavity $Q$-factor is also reduced. This subsequent loss trivially lowers the finesse ($\mathcal{F}$) of the cavity, hence only slight changes in the $Q$-factor are introduced despite the higher effective index ($\kappa_{eff}$) change, since increase in mirror quality with tuning (higher reflection since more metallic) is compensated by the higher loss. For our ITO slot PCNB cavity we achieve the $Q$ value of ~774 (695) in the ON (OFF) state, where the lower $Q$-value of the OFF state originates from the modal absorption increase (Fig. 3d). The $Q$-factors calculated here use a low-$Q$ cavity method used in the Lumerical FDTD solutions, which is determined through the Fourier transform of the field by finding the resonant frequencies ($f_R$) of the signal and measuring the full width half maximum (FWHM, $\Delta f$) of the resonant peaks, i.e., $Q = f_R/\Delta f$, if the electromagnetic fields decay completely from the simulation in a time that can be reasonably simulated by FDTD [28].

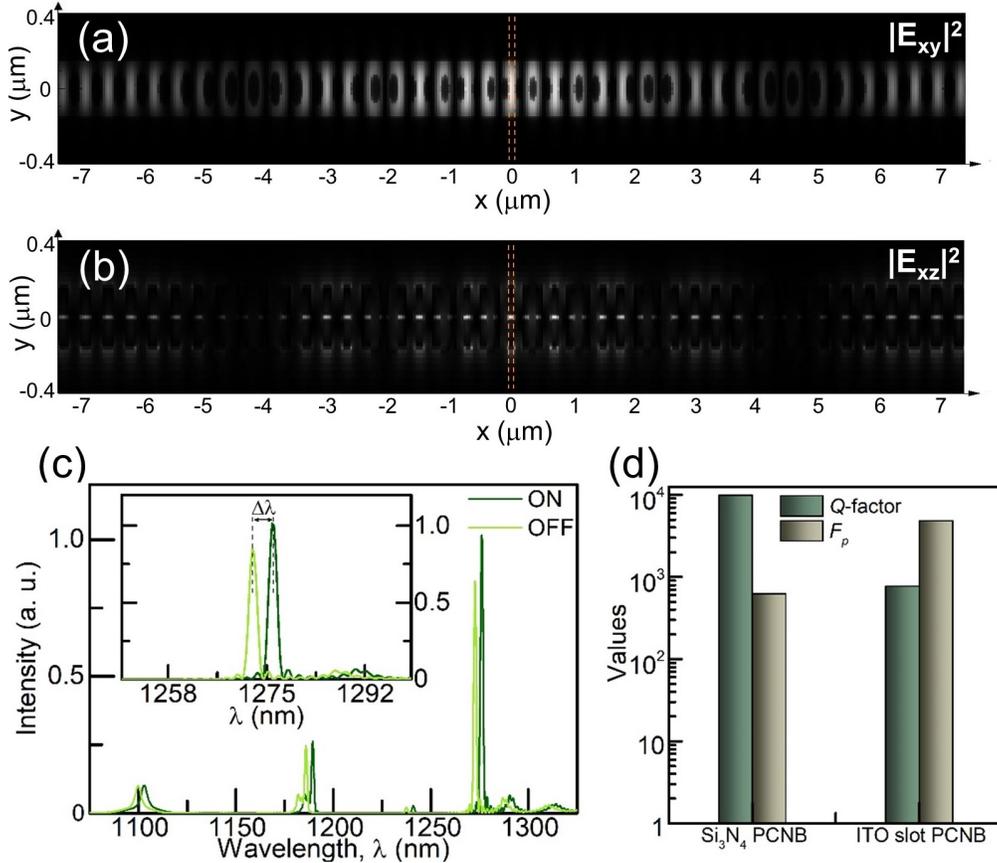

**Figure 3:** Electric field intensity, $|\mathbf{E}|^2$, distribution profiles of the cavity mode. (a) TM-like 1$^{st}$ order mode of 1D PCNB cavity, and $|\mathbf{E_{XY}}|^2$ is recorded in the xy-plane at the cavity center along the z-direction. (b) $|\mathbf{E_{XZ}}|^2$ is recorded in the xz-plane along the y = 0 direction. Both the cavity field profiles are recorded at the middle of the active layer height at z = 0.16 $\mu m$, and the cavity region is centered at x = 0 in the x-direction marked with orange dashed lines. (c) Transmission intensity vs. wavelength, $\lambda$ corresponding to states of operation; the tunability of the cavity is shown in the inset, $\Delta\lambda \sim 3.4\ nm$. (d) The quality factor, $Q$ and Purcell factor, $F_p$ of the ITO vertical slot PCNB cavity compared to a similar Si$_3$N$_4$ 1D PCNB cavity [14].

The longitudinal (z-axis) confinement leads to the necessary cavity feedback to facilitate stimulated emission (lasing), if optically or electrically pumping the gain medium. In recent history, the Purcell factor, $F_p$, is used to describe the LMI in laser physics as it relates to a measure of the spontaneous emission rate enhancement of a dipole emitter source placed in the cavity compared to internal radiative recombination rates in a homogeneous semiconductor material, given by [29]

$$F_p = \frac{3}{4\pi^2}\left(\frac{\lambda_R}{n}\right)^3 \left(\frac{Q}{V_{mode}}\right) \quad (2)$$

where $\lambda_R$ is the resonant free-space wavelength of the cavity, $n$ is the real part of the complex refractive index at the field antinode, and $V_{mode}$ can be estimated from a commonly used definition as [14]

$$V_{mode} = \frac{\int \epsilon |E(r)|^2 dV}{\max\{\epsilon |E(r)|^2\}} \qquad (3)$$

where $\epsilon$ is the dielectric constant, $E(r)$ is the electric field strength, and $V$ is a quantization volume encompassing the resonator and with a boundary in the radiation zone of the cavity. Eq. (2) indicates that a large $Q$ and a smaller $V_{mode}$ are desired to enhance emission rate (i.e. $F_p$). Evidently, there exist two techniques to increase the Purcell factor – the classical approach is enhancing the cavity $Q$-factor, and the other one is minimizing the mode volume, $V_{mode}$. However, the former is rather unpractical since it not only requires increased wafer space, but also has negative effects for data communication applications, particularly for lasers driven in a direct modulation mode due to the long cavity photon lifetimes inside the high-$Q$ cavity, $\tau_{ph} = Q\lambda/2\pi c$, where $\lambda$ is the operating resonant wavelength and $c$ is the speed of light in vacuum. Although the small mode volumes of PCNB cavities can be attained by a proper design, high-$Q$ factors are typically obtained using extensive parameter search and optimization [30]. As our slot structure can squeeze in the light in the active region due to higher confinement arising from the index contrast, the small modal area leads to a miniscule diffraction limited mode volume in a cubic half-wavelength in material of ~0.1 $(\lambda/2n)^3$ in the cavity. Since $Q$ is ultimately limited in practice by other factors, such as bandwidth considerations, material absorption, or fabrication tolerances, minimizing $V_{mode}$ for a given $Q$ is a preferred solution with practical applications in mind. Such optical cavities potentially enable spontaneous emission rates that are faster than the stimulated emission rates for nanoscale light-emitting devices [31, 32]. Note, this is not only limited to nanoscale-based laser cavities but also can be applied to conventional laser designs as well [14]. On the other hand, the internal dynamics leading toward the laser threshold are more efficiently utilized when a smaller optical mode volume is used (i.e., higher $F_p$ and spontaneous emission coupling factor, $\beta$). The case with a smaller mode volume often translates into a low power requirement. It is worthy to mention that while traditional techniques involving tunable lasers utilize changing the feedback system of the cavity usually by changing the cavity length (i.e. effective path length), our proposed design can utilize the tunability from effective index tuning in the cavity free from moving parts and providing degrees of freedom in design.

Comparing the cavity performance of a $Si_3N_4$ 1D PCNB cavity from [14] to our results shows that the $Q$-factor has declined by about one order of magnitude because of the addition of the lossy ITO, yet the Purcell factor has risen by about the same amount (Fig. 3d); in detail, $Q$-factor ~10,000 for the $Si_3N_4$ PCNB to ~770 for the ITO slot PCNB, which can be contributed to the inherent orders of magnitude higher loss in the ITO material. However, we observe a significantly higher Purcell factor enhancement from ~620 for the $Si_3N_4$ PCNB to ~4820 (~4320) for the ITO slot PCNB cavity ON (OFF) state. This can be attributed to the strong modal confinement in our vertical slot structure, essentially squeezing the light into the active region arising from sufficient index contrast in the mode. As such, we are able to reduce the effective modal area resulting in a modest small mode volume.

## 5. Conclusion

In conclusion, we have shown that one-dimensional photonic crystal nanobeam cavity can be used to realize a highly wavelength-shiftable device. Such functionality is based on exploiting the strong index tunability of ITO, the sub-diffraction limited optical confinement of the vertical photonic (non-plasmonic) slot waveguide design, and the cavity feedback. Taken together, we find a strong light matter interaction of such a system indicative of a Purcell factor exceeding ~4000, yet only requiring a moderate $Q$-factor of about 770. We find a wide tuning range of ~3.4 nm in the telecom O-band with only about 10% $Q$-factor suppression. Tuning originates from the carrier concentration altering of ITO through bias voltages. The demonstrated enhancement in electro-optic tunability enables compact, power-efficient, and reconfigurable integrated photonic components, such as on-chip light sources including tunable lasers and optical modulators for future photonic circuitry.

## Acknowledgements

V.S. is supported by ARO (W911NF-16-2-0194), by AFOSR (FA9550-17-1-0377). H.D. and V.S. are supported by AFOSR (FA9550-17-P-0014) of the small business innovation research (SBIR) program.